\documentclass[twocolumn,showpacs,preprintnumbers,amsmath,amssymb,prb]{revtex4}

\usepackage{graphicx}
\usepackage{dcolumn}
\usepackage{bm}

\begin{document}

\preprint{APS/123-QED}

\title{A novel magnetic phase prior to a saturation moment in frustrated spinel oxides in ultra-high magnetic fields}

\author{Atsuhiko Miyata}
 \affiliation{Institute for Solid State Physics, The University of Tokyo, 5-1-5, Kashiwanoha, Kashiwa, Chiba, 277-8581, Japan}
 
\author{Hiroaki Ueda}
 \affiliation{Department of Chemistry, Graduate School of Science, Kyoto University, Kyoto, 606-8502, Japan}
  
\author{Shojiro Takeyama}
\email{takeyama@issp.u-tokyo.ac.jp}
 \affiliation{Institute for Solid State Physics, The University of Tokyo, 5-1-5, Kashiwanoha, Kashiwa, Chiba, 277-8581, Japan}

\date{\today}

\begin{abstract}

We have investigated the magnetic-field induced phases of a typical three-dimensional frustrated magnet, CdCr$_2$O$_4$, in magnetic fields of up to 120 T that is generated by the single-turn coil techniques. 
We focused on magnetic phase transitions in proximity of a saturated magnetization moment.
We utilized both the electromagnetic induction method using magnetic pick-up coils and magneto-optical spectroscopies of the $d$-$d$ transitions and the exciton-magnon-phonon transitions to study the magnetic properties subjected to ultra-high magnetic fields.
Anomalies were observed in magneto-optical absorption intensity as well as differential magnetization prior to a fully polarized magnetic phase (a vacuum state in the magnon picture), revealing a novel magnetic phase associated with changes in both crystal and magnetic structures accompanied by the first order phase transition. 
Magnetic superfluid state such as an umbrella-like magnetic structure or a spin nematic state, is proposed as a candidate for the novel magnetic phase, which is found universal in the series of chromium spinel oxides, $A$Cr$_2$O$_4$ ($A$ = Zn, Cd, Hg).

\end{abstract}


\maketitle

\section{Introduction}
Application of magnetic fields to geometrically frustrated magnets induces a variety of novel magnetic phases, such as the magnetization plateau phase~\cite{okamoto11} and the supersolid phase~\cite{miyataJPSJ11}. 
These phases emerge as a result of the interplay between the magnetic field and the competing magnetic interactions. 
This has prompted intensive research aimed at exploring novel magnetic phases that are peculiar to geometrically frustrated magnets in recent decades. 
As for the vicinity of a fully polarized phase, theoretical works predict the appearance of novel magnetic phases such as the Bose-Einstein condensation of two-magnon bound states (a spin nematic phase)~\cite{zhitomirsky10} and a crystallization of magnons~\cite{schulenburg02, zhitomirsky04} based on a magnon picture in which the fully polarized state is regarded as a vacuum state. 
Yet, relatively few experimental studies have been conducted so far because extreme physical conditions such as ultrahigh magnetic fields are of necessity for experimental investigation of magnetic phases in the geometrically frustrated magnets up to full magnetization moment because of their extremely strong antiferromagnetic coupling. 
Therefore, the development of techniques that allow measurements to be performed in ultra-high magnetic fields has played a crucial role in elucidation of physics that originate from the geometrical frustration. 

Until recently, the Faraday rotation measurement, from which a magnetization curve is reproduced under certain conditions ~\cite{nojiriPRB95}, has been almost the only method available to investigate magnetic properties under ultra-high magnetic fields (above 100 T)~\cite{nojiriJPCM95, kojima08}. 
However, it was difficult to identify all magnetic phases only from the magnetization curves. 
For example, in the case of ZnCr$_2$S$_4$, which is considered a bond-frustrated magnet~\cite{rudolf07}, ultrasound measurements showed a discernible anomaly. 
While this anomaly indicated the existence of phase transition that was accompanied by the change of crystal structure, the magnetization curve monotonically increased up to a saturation moment, without any noticeable anomalies~\cite{tsurkan11}. 
This example demonstrates that the magnetization curve is not always sufficient to reflect the magnetic phases.

Magnetization measurement systems have been substantially improved recently by utilizing the electro-magnetic induction method adjusted for use in a single-turn coil method capable of measuring up to magnetic fields of above 100 T~\cite{takeyama11}. 
Consequently, the sensitivity of magnetization changes became much higher than that obtained using the Faraday rotation measurement method in this range of high magnetic fields. 
Furthermore, absorption spectral measurements on the $d$-$d$ transition and exciton-magnon-phonon (EMP) transition were shown to be a powerful tool that allow the detection of any changes in crystal as well as magnetic structures in magnetic fields~\cite{miyataPRL11, miyata12}. 
This method was shown to be as useful as the ultrasound and magnetostriction measurements~\cite{bha11, jaime11}. 
Moreover, the absorption spectral measurements can derive information related to magnetic structures from the EMP transitions, in addition to that of crystal fields.

In chromium spinel oxides, $A$Cr$_{2}$O$_{4}$ ($A =$ Zn, Cd, Hg), Cr$^{3+}$ ions form a pyrochlore lattice. 
Therefore, these compounds have been recognized as typical three-dimensional geometrically frustrated magnets~\cite{rudolf07}. 
The magnetic properties of chromium spinel oxides strongly depend on the spin-lattice interactions~\cite{rudolf07, kant09}. 
Complicated magnetic structures associated with lattice distortions appear at the N\'eel temperature of chromium spinel oxides; this phenomenon was termed ``order by distortion''~\cite{chung05, ji09}. 
When subjected to a strong magnetic field, these compounds exhibit rich magnetic phases such as the robust half-magnetization plateau phase~\cite{kojima08, ueda06}. 
These phases were well captured by a bilinear-biquadratic model that considers spin-lattice interactions, as was first theoretically described by Penc \textit{et al.}~\cite{penc04}. 
In addition, we were the first to report a canted 2:1:1 phase that preceded the half-magnetization plateau phase in ZnCr$_{2}$O$_{4}$~\cite{miyataJPSJ11}, which was already predicted theoretically for a small limit of the biquadratic term in the model~\cite{penc04}. 
On the other hand, magneto-optical absorption spectral measurements, described above, have revealed a novel magnetic phase prior to a fully polarized phase in ZnCr$_{2}$O$_{4}$~\cite{miyataPRL11, miyata12}. 
This observation was beyond the framework of the existing theory that is based on the bilinear-biquadratic model~\cite{penc04}. 
This novel phase was known to be very important in a boson picture, as it corresponds to a superfluid state sandwiched between the supersolid and the liquid states, in good analogy with the theory that was developed for $^4$He quantum phases~\cite{miyataPRL11, miyata12}. 
This prompted the question of whether the novel phase that precedes the saturation moment is observable only in ZnCr$_{2}$O$_{4}$ or whether it exists in other chromium spinel oxides that have different strengths of spin-lattice interactions. 

Magnetization curves up to a full saturation moment have already been obtained for CdCr$_{2}$O$_{4}$ using the Faraday rotation measurement method in magnetic fields of up to 140 T generated by the single-turn coil technique, but with a limited resolution~\cite{kojima08}. 
Here, we focus on high-magnetic-field phases of CdCr$_{2}$O$_{4}$ (N\'eel temperature $T_\text{N} = 7.8$ K and Curie-Weiss temperature $\theta_\text{CW} = -70$ K~\cite{ueda05}), investigated in more detail using both magnetization and magneto-optical absorption spectral measurements that are conducted in ultra-high magnetic fields.

\section{Experimental procedure}

We conducted the magnetization measurements using the induction method under ultrahigh magnetic fields (above 100 T) using a vertical single-turn coil (VSTC) method that has been developed recently and is discussed in detail in Ref.~\cite{takeyama11}. 
In addition, magneto-optical absorption spectral measurements were performed using a streak camera; these measurements were conducted using a horizontal single-turn coil (HSTC) method, and the optical system was set up in a manner similar to that described in Ref.~\cite{miyata12}. 
In both measurements, magnetic fields of up to 120 T were generated. 
To achieve low temperatures, we have used a hand-made liquid-He container made of glass-epoxy for the VSTC method~\cite{takeyama11} and a hand-made liquid-He flow-type cryostat made totally of Stycast (No. 1266) resin for the HSTC method~\cite{miyata12}. 
The sample temperature was monitored by an Au-Fe/Chromel thermocouple attached adjacent to the sample. 
In the case of HSTC, the accuracy was $\pm$1 K, whereas in the case of VSTC, the accuracy was less than 0.1 K. 
A single crystal of CdCr$_2$O$_4$ was grown using the flux method, cut parallel to the (111) crystal surface, and attached on a quartz substrate for magneto-optical measurements; the diameter and thickness of the sample were $\sim$2 mm and 50 $\mu$m, respectively. 
A powdered sample of CdCr$_2$O$_4$ was used for magnetization measurements; the net mass was 5.4 mg. 
The value of magnetic field was measured using a calibrated pick-up coil wound near the sample. 
The estimated error of the value of the magnetic field was less than 3$\%$. 

\section{Results and Discussion}
\subsection{Magnetization measurements}
 
Figure 1 shows both the derivative, $dM/dH$, of the magnetization curve, and its integrated $M$ curve, obtained for CdCr$_{2}$O$_{4}$ at 4.2 K in magnetic fields up to 120 T by the induction method using a magnetic pickup coil~\cite{takeyama11}. 
The full-magnetization curve is in fairly good agreement with those obtained previously using the Faraday rotation measurement method~\cite{kojima08}. 
The $dM/dH$ curve showed sharp peaks at around 28 T as well as a hysteresis in elevating and descending processes of the magnetic field. 
This indicates a first-order phase transition between the antiferromagnetic phase and the half-magnetization plateau phase. 
This phase transition was reported earlier by Ueda \textit{et al.}~\cite{ueda05} based on magnetization measurements of CdCr$_{2}$O$_{4}$ in magnetic fields of up to 50 T. 
At around 58 T, we observed a step-like structure in the $dM/dH$ curve. 
As described in Ref.~\cite{kojima08, bha11}, this step-like structure indicates a second-order phase transition between the half-magnetization plateau phase and a canted 3:1 phase. 
In addition, for the $dM/dH$ curve, we observed peaks at 74 T (for descending magnetic field) and 77 T (for elevating process), indicating the existence of a first-order phase transition. Although this transition was observed earlier by Mitamura \textit{et al.}~\cite{mitamura07}, only the descending magnetic field process of the $dM/dH$ curve from 80 T was shown in their data, because owing to a large starting trigger noise, it was difficult to analyze the signal in the elevating magnetic field process. 
A new observation in the present study is a step-like structure that appears in the $dM/dH$ curve at around 88 T, which should be considered the phase transition to the fully polarized phase. 
The present data showed the existence of a new magnetic phase between 74 -- 77 T and 88 T. 
This new phase cannot be explained within the existing framework of the bilinear-biquadratic model developed by Penc \textit{et al.}~\cite{penc04} that incorporates spin-lattice interactions because this model cannot allow any phase between the canted 3:1 phase and the fully polarized phase.

\begin{figure}
    \centering
    \hfill
    \includegraphics[width=0.45\textwidth]{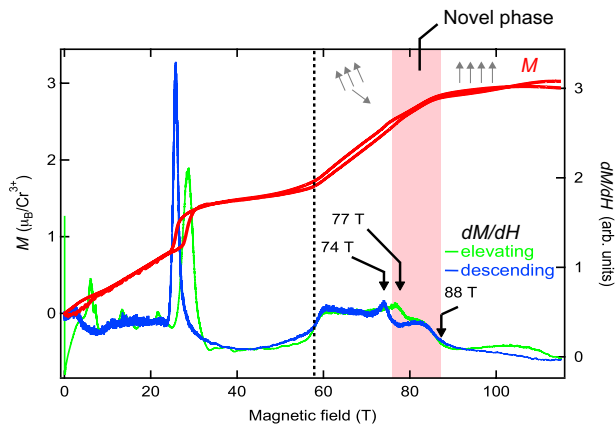}
    \hfill\null
\caption{\label{fig:dmdh} 
(Color online) Magnetization and $dM/dH$ curves of CdCr$_2$O$_4$ at 4.2 K in magnetic fields up to 120 T obtained by the electro-magnetic induction method. 
Black arrows show anomalies observed in the vicinity of the fully polarized magnetic phase. Gray arrows illustrate a canted 3:1 and a fully polarized magnetic structure.
}
\end{figure}

\subsection{Magneto-optical absorption spectra}

To confirm the existence of the novel phase between 74 -- 77 T and 88 T in CdCr$_2$O$_4$, we performed magneto-optical absorption spectral measurements.
Our recent magneto-optical absorption studies conducted for ZnCr$_2$O$_4$ have revealed a novel magnetic phase between a canted 3:1 phase and the fully polarized phase~\cite{miyataPRL11, miyata12}.

Figure 2 shows magneto-optical absorption spectra of CdCr$_2$O$_4$ obtained at 6 K for several values of magnetic field. 
In Fig. 2, both $d$-$d$ ($^4A_2~\rightarrow~^4T_2$) transition and the exciton-magnon-phonon (EMP) transition~\cite{kojima00, szm80} are observed. 
Because these transitions reflect the crystal field and magnon excitation, the spectral shape and peak position are susceptible to the change of crystal and magnetic structure~\cite{kojima00, eremenko76}. 
The peaks of EMP transition lose their intensity for increasing magnetic field and are completely suppressed at 100 T because a magnon with $\Delta S^z = +1$ cannot be excited in the fully polarized phase~\cite{eremenko76}.

\begin{figure}
    \centering
    \hfill
    \includegraphics[width=0.45\textwidth]{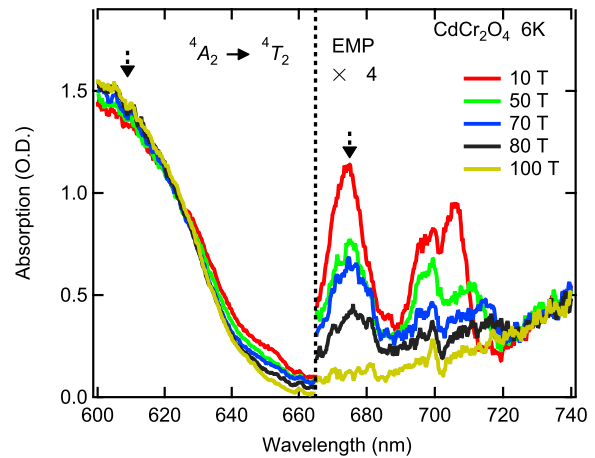}
    \hfill\null
\caption{\label{fig:spectra} 
(Color online) Magneto-optical absorption spectra obtained at 6 K in a region of wavelengths where $^4A_2~\rightarrow~^4T_2$ and EMP transitions occur in CdCr$_2$O$_4$. The spectra were measured at several magnetic fields (up to 100 T) obtained by streak spectroscopy. Absorption intensity of EMP transitions is multiplied by four. Dotted arrows show wavelengths at which the absorption intensities in magnetic fields are plotted in Fig. 3.
}
\end{figure}

In Fig. 3, we plotted the magneto-optical absorption intensity at selected wavelengths of 610 nm ($^4A_2~\rightarrow~^4T_2$ transition) and 675 nm (EMP transition) as a function of magnetic field. 
For these transitions, the first anomaly in absorption intensity was observed at 28 T, where the first-order phase transition occurred, being accompanied by change of magnetic and crystal structures as described in Ref.~\cite{matsuda10}. 
Furthermore, for the EMP transition below 28 T, there is a clearly observable large difference between the absorption intensities of the elevating and descending magnetic field processes. 
This might indicate that for magnetic fields below 28 T, magnetic and/or crystal structures were different between the two processes. 
One possible explanation of this behavior is that the magnetic domain orientation and fast sweep rate of magnetic fields ($\sim$60 T/$\mu$s) caused the large hysteresis behavior. In fact, neutron scattering measurements of CdCr$_2$O$_4$ in fields of up to 30 T exhibited hysteresis behavior that arose from magnetic domain orientation at $\sim$8 T (sweep rate in those experiments was $\sim$4000--5000 times slower than in our experiments)~\cite{matsuda10}. 
An additional indication of the change of crystal and magnetic structures are anomalies of absorption intensity at 74 T in these transitions, corresponding to the first-order phase transition that is observed in Fig. 1. 
We consider that this change of crystal structure occurred to relieve the lattice distortion that emerged from the geometrical frustration and turned into a higher symmetric structure. 
In addition, a change of the lattice structure would persist up to the magnetic field of 88 T, because absorption intensity in the $^4A_2~\rightarrow~^4T_2$ transition (which reflects the crystal field) exhibits a monotonic increase up to 88 T, until it becomes almost constant. 
On the other hand, for EMP transition, the absorption intensity was completely suppressed above 88 T, because excitation of a magnon with $\Delta S^z = +1$ is substantially restricted in the fully polarized phase~\cite{eremenko76}. 
These behaviors of absorption intensity in CdCr$_2$O$_4$ are very similar to those observed in the vicinity of the fully polarized phase in ZnCr$_2$O$_4$, indicating the existence of the novel phase in both species of chromium spinel oxides.

\begin{figure}
    \centering
    \hfill
    \includegraphics[width=0.4\textwidth]{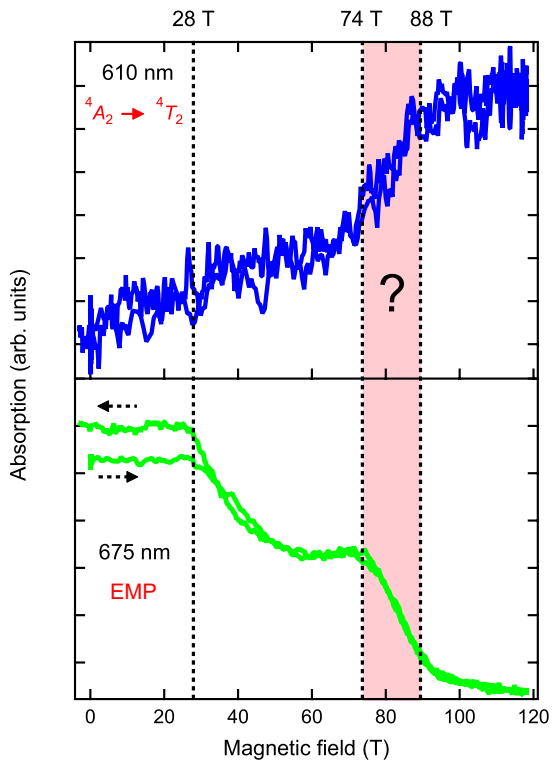}
    \hfill\null
\caption{\label{fig:absorption} 
(Color online) Magneto-optical absorption intensity at selected wavelengths of 610 nm ($^4A_2~\rightarrow~^4T_2$ transition) and 675 nm (EMP transition) plotted as a function of magnetic field. 
Dotted lines show anomalies observed in the absorption intensity.
}
\end{figure}

\subsection{Novel magnetic phase in Cr spinel oxides}

Magneto-optical absorption spectral and $dM/dH$ measurements could indicate evidence of the novel phase in CdCr$_2$O$_4$. 
Results of these measurements enabled us to compare, for the first time, novel behaviors that are observed in the series of chromium spinel oxides, CdCr$_2$O$_4$, HgCr$_2$O$_4$, and ZnCr$_2$O$_4$ prior to the fully polarized phase. 
In the case of HgCr$_2$O$_4$, of which the saturated magnetization moment is reached for much lower magnetic fields, it has been difficult to conduct magneto-optical absorption spectral measurements, because only polycrystalline samples are available. 
However, the behavior of the $dM/dH$ curve was very similar to that observed in the present case~\cite{kimura11}. 
Recently, Kimura \textit{et al.} have reported the possibility that the novel magnetic phase may exist prior to the fully polarized phase in HgCr$_2$O$_4$, where a relief of lattice distortions occurred; these observations were based on the electron spin resonance measurements~\cite{kimura11}. 
This report also supported the idea that the lattice symmetry in the novel phase becomes very high. Similar behaviors are observed in the absorption intensity and/or in the $dM/dH$ prior to the full saturation, suggesting that the novel phase exists universally in different species of chromium spinel oxides. 
Figure 4 shows a magnetic phase diagram of chromium spinel oxides, for which their antiferromagnetic interactions $J$ were derived from the Curie-Weiss temperature (-390 K, - 70 K, and -32 K in ZnCr$_2$O$_4$, CdCr$_2$O$_4$, and HgCr$_2$O$_4$, respectively~\cite{ueda06}) and a Land\'e $g$-factor of $g =$ 2, in a manner similar to what has been described in Ref.~\cite{miyata12}. 
The values of the phase transition points were taken from Ref.~\cite{miyata12} and Ref.~\cite{kimura11} for ZnCr$_2$O$_4$ and HgCr$_2$O$_4$, respectively. 
According to the Ref.~\cite{penc04, bergman06}, the strength of spin-lattice coupling $\alpha$ is defined by the relation $J\alpha \sim -dJ/dr$, where $r$ is distance between two relevant spins. 
Since there obtained almost linear relationship between the Curie-Weiss temperature (the strength of the antiferromagnetic interaction) and the lattice constant in these chromium spinel oxides~\cite{rudolf07}, it could be suggested that $dJ/dr$  is approximately constant at the vicinity of their lattice constant. 
In fact, $dJ/dr$ in HgCr$_2$O$_4$ and CdCr$_2$O$_4$ have been found to take almost the same value by the measurements of the magnetic properties under pressure~\cite{ueda08}. 
In Fig. 4, we use the relation $J\alpha \sim 1$ for simplicity. The plot shows that the existence of the novel phase does not depend on the strength of spin-lattice coupling.

\begin{figure}
    \centering
    \hfill
    \includegraphics[width=0.4\textwidth]{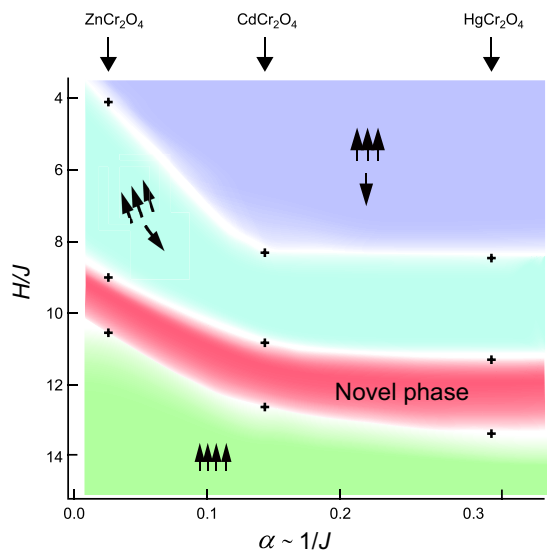}
    \hfill\null
\caption{\label{fig:diagram} 
(Color online) Magnetic phase diagram of chromium spinel oxides.
The phase transition points were taken from Ref.~\cite{miyata12} for ZnCr$_2$O$_4$ and from Ref.~\cite{kimura11} for HgCr$_2$O$_4$.
}
\end{figure}

\subsection{Possible origin of novel magnetic phase}

A magnon (bosonic) picture is more useful for elucidation of the physics underlying at the vicinity of a saturated moment.  
A spin system is compared with a boson system such as $^4$He, and the discussion was developed in the similar system of ZnCr$_2$O$_4$~\cite{miyata12}. 
$S^z$ components of spins break the translational symmetry in a magnetic ``solid'' state, and their $S^x$ and $S^y$ components break spin rotational symmetry in a magnetic ``superfluid'' state~\cite{matsuda70, liu73}. 
A magnetic ``supersolid'' state has both properties, whereas the translational and spin rotational symmetries are both maintained in a magnetic ``liquid'' state~\cite{matsuda70, liu73}.  
In general, the supersolid state exists between the solid and the superfluid states in the phase diagram, because a direct transition from a liquid state to a supersolid state or vice versa require simultaneous breaking of the translational and the spin rotational symmetries, which could occur with less possibility. 
Therefore, the novel magnetic phase, which is located between a supersolid state (canted 3:1 phase) and a liquid state (fully polarized phase), should be a superfluid state as was discussed in Ref.~\cite{miyata12}. 
Furthermore, present experimental results reveled that the magnetic structure and the crystal structure become highly symmetric in the novel phase under discussion. 
As for the concrete spin structure corresponding to the magnetic superfluid state, there are two realistic candidates, one is an umbrella-like magnetic structure and the other is a spin nematic state. 
The magnetic structures of both candidates are highly symmetric, and break the spin rotational symmetries. 

Interestingly, the existence of the umbrella-like magnetic structure prior to the fully polarized phase was also discussed for the case of CuFeO$_2$, which is regarded as a geometrically frustrated magnet with a triangular lattice~\cite{lummen10}. 
In this compound, as well as in chromium spinel oxides, a biquadratic term that arises from the spin-lattice interactions plays an important role in the stabilization of several collinear structures~\cite{lummen10}. 
When a magnetic field was applied perpendicular to the $c$-axis in CuFeO$_2$, a sequence of magnetic phases appeared, similar to what has been observed for chromium spinel oxides~\cite{lummen10}. 
However, the complicated magnetic phase diagram of CuFeO$_2$ cannot be explained by the model that includes the biquadratic term~\cite{lummen10}. 
Therefore, Lummen \textit{et al.}~\cite{lummen10} suggested that additional contributions such as longer-range biquadratic interactions and quantum effects should be considered for the accurate description of CuFeO$_2$ magnetism. 
Similarly, these additional contributions might be crucially important in understanding the phase diagram of chromium spinel oxides. 
These compounds might have a common underlying mechanism to stabilize the umbrella-like magnetic structure.

Thus, one way to explain the stabilization of the umbrella phase is to consider the additional classical effects that favor noncollinear magnetic structures. 
As described in Ref.~\cite{penc07}, if a coefficient of the biquadratic term changes its sign, noncollinear magnetic structures will be favored and the umbrella phase will stabilize prior to the fully polarized phase. 
However, in the bilinear-biquadratic model proposed by Penc \textit{et al.}~\cite{penc04}, the sign of the coefficient is not likely to change. 
Therefore, additional terms should be incorporated into the bilinear-biquadratic model. 
In fact, a magnetic structure that is observed in the half-magnetization plateau phase of CdCr$_2$O$_4$~\cite{matsuda10} indicates the necessity of additional terms such as three-site spin-lattice interactions suggested by Bergman \textit{et al.}~\cite{bergman06}. 
Such interaction makes the magnetic structure less collinear. 
Therefore, if the contributions from the additional terms such as the above-described interaction become more dominant than those from the biquadratic terms, the umbrella phase might exist prior to the fully polarized phase. 

Another possibility is to consider quantum effects. 
A magnon picture would be useful to understand the magnetic properties in the vicinity of a saturated moment.
If the magnetic field is sufficiently high, the fully polarized phase becomes the ground state. 
A one-magnon state is represented as the state in which the quantum number $S^z$ is lowered by 1. 
In the umbrella phase, magnons are uniformly distributed throughout the lattice, whereas the canted 3:1 phase is characterized by partial magnon localization. 
The energy difference between both states depends on interactions between magnons. 
In chromium spinel oxides, quantum effects might cause the interaction that favors the umbrella phase prior to the fully polarized phase. 

From this viewpoint, a two-magnon bound state Bose-Einstein condensation, a spin nematic phase, could occur in the vicinity of a fully polarized phase under certain conditions of magnon-magnon interactions~\cite{zhitomirsky10}. 
Experimental observations of the spin nematic phase prior to a fully polarized phase were recently reported for the frustrated chain compound LiCuVO$_4$, in which the nearest neighbor ferromagnetic and the next nearest neighbor antiferromagnetic interactions stabilize the spin nematic phase~\cite{svistov11}. 
Stabilization of the spin nematic phase in bilinear-biquadratic model with quantum effects was suggested earlier~\cite{chubukov90, lauchli06}. 
L\"auchli \textit{et al.} reported that the spin nematic phase could exist prior to the fully polarized phase in the $S$ = 1 bilinear-biquadratic Heisenberg model on a triangular lattice~\cite{chubukov90, lauchli06}. 
Similarly, there would be no reason to exclude the possibility of the spin nematic phase in chromium spinel oxides that are well described by the bilinear-biquadratic Heisenberg model on pyrochlore lattice.

A more sophisticated model that accounts for the higher-order spin-lattice interactions, quantum effects, and so forth is required to clarify the magnetic structure of the novel phase universally observed in the series of chromium spinel oxides.

\section{Conclusions}

The magneto-optical absorption spectral measurements and magnetization measurements were conducted on a typical three-dimensional frustrated magnet, CdCr$_2$O$_4$, in ultra-high magnetic fields of up to 120 T using the induction method. 
As a result, the anomalies in magneto-optical absorption intensity in the $^4A_2~\rightarrow~^4T_2$ and the EMP transitions were observed prior to the fully polarized phase. 
These anomalies indicate the existence of the novel magnetic phase that has been hardly resolved using only the Faraday rotation measurement method. 
In addition, the $dM/dH$ curve also showed a distinct peak associated with a hysteresis at the transition. 
These behaviors were similar to those observed for the absorption intensity of ZnCr$_2$O$_4$ and for the $dM/dH$ curve in HgCr$_2$O$_4$, thus indicating that the umbrella phase (or other phase, such as the spin nematic phase) should exist prior to full saturation, commonly in chromium spinel oxides. 
To understand the novel phase, we should consider a more sophisticated model that includes additional effects such as higher-order spin-lattice interactions and quantum effects.

\section*{Acknowledgments}

We would like to acknowledge Y. Motome, N. Shannon, K. Penc, M. Oshikawa, and Y. Chen for fruitful discussions.

\end{document}